**POINT OF VIEW**

Running head: MORE TAXA ARE NOT NECESSARILY BETTER

**More Taxa Are Not Necessarily Better
for the Reconstruction of Ancestral Character States**

GUOLIANG LI[1], MIKE STEEL[2] and LOUXIN ZHANG[3]

[1] *Department of Computer Science, National University of Singapore, Singapore 117543;*
ligl@comp.nus.edu.sg
[2] *Biomathematics Research Centre, University of Canterbury, Christchurch, New Zealand;*
M.Steel@math.canterbury.ac.nz
[3] *Department of Mathematics, National University of Singapore, Singapore 117543;*
matzlx@nus.edu.sg

Communicating author:
  LX Zhang; Phone: +65-6516-6579; Fax: +65-67795452;
  Email: matzlx@nus.edu.sg.

Ancestral state reconstruction is an important approach to understanding the origins and evolution of key features of different living organisms (Liberles, 2007). For example, ancestral proteins and genomic sequences have been reconstructed for understanding the origins of genes and proteins (Blanchette et al., 2004; Cai, Pei and Grishin, 2004; Gaucher et al., 2003; Hillis et al., 1994; Jermann et al., 1995; Taubenberger et al., 2005; Thornton et al., 2003; Zhang and Rosenberg, 2002). A variety of reconstruction methods including parsimony and maximum likelihood exist for bio-molecular sequencing (Yang et al., 1995; Elias and Tuller, 2007; Koshi and Goldstein, 1996), multistate discrete data (Pagel, 1999, Schultz et al., 1996; Mooers and Schluter, 1999) and continuous data (Martins, 1999). These different reconstruction methods have been assessed by both theoretical analyses (Maddison, 1995; Elias and Tuller, 2007) and computer simulation (Blanchette et al., 2004; Moorers, 2004; Salisbury and Kim, 2001, Schultz et al., 1996; Williams et al., 2006; Zhang and Nei, 1997). One important observation in these investigations is that the topology of the phylogenetic tree relating the extant taxa to the target ancestor has a significant influence on reconstruction accuracy. For instance, a star-like phylogeny allows the ancestral character states to be inferred more accurately than other topologies given the same number of terminal taxa under the two-state symmetric model (Evans et al., 2000; Schultz et al., 1996). For more complex models (e.g. on four-states such as DNA), the influence of topology on reconstruction accuracy is more complicated (Lucena and Haussler, 2005).

**MODELS AND METHODS**

We study how ancestral state reconstruction depends on taxon sampling, with the assumption that the true phylogenetic tree is given. Intuitively, more terminal taxa should give better reconstruction accuracy. For example, in a recent review, Crisp and Cook (2005) recommend that "if ancestral features are to be inferred from a phylogeny, a method that optimizes character states over the whole tree should be used." In certain cases, this viewpoint can be formally justified; for example, consider the problem of estimating the root state given a character at the leaves of a tree, under a model of character evolution in which the branch lengths are known, and each of the states at the root has equal prior probability. In this case, the 'most accurate' method for reconstructing the root state is to use a local (or marginal) maximum likelihood (ML) method, applying it to the total set of taxa (not just a subset). Before we justify this claim, recall that for estimating the root state, the local ML procedure simply selects the state that has the highest probability of evolving the given characters under the model (with the branch lengths specified), and any ties are broken uniformly (Koshi and Goldstein, 1996; Schluter et al, 1997; Felsenstein, 2004). By 'most accurate', we mean the method that has the highest expected probability of returning the correct root state. The proof of the claimed optimality of ML in this setting can be found in Berger (1985, p.159), or Steel and Szekely (1999, Theorem 4). For models in which certain states may have higher *a priori* probability at the root, the most accurate reconstruction method is to maximize the posterior probability of the root state (i.e. the product of the ML score for each root state with its *a priori* probability). In summary, if the branch lengths and model are known, it is always best to use all the terminal taxa, and to do so in an ML-style framework (which can provide a different estimate from that provided by maximum parsimony).

However, if the branch lengths that describe the evolution of the character are not known, the situation is more complicated. For simple models, such as the symmetric Poisson model (e.g. the Jukes-Cantor model on four states), it is known that the ML estimate of the root state (where one also optimizes the branch lengths as 'nuisance' parameters) is identical to the maximum parsimony estimate (see Theorem 6, Tuffley and Steel, 1997).

This leads to a natural question regarding the situation where the branch lengths for the character are unknown to the investigator – is the maximum parsimony estimation of the root state using all the leaf species more accurate than just using a subset of leaf species? We will see that for certain trees, it may be better to use some leaves – or even a single leaf – that is 'near' the root for estimating the root state.

Given a phylogenetic tree of a group of taxa, we assume that the character evolves by a Markov process, starting with a state at the root and proceeding to the leaves. The evolutionary model specifies the length of each branch or, equivalently, the probability that a state $c$ evolves to a state $d$ on a branch from node $f$ to node $x$ as conditional probability $\Pr[s_x = d \mid s_f = c]$.

In this study, we also assume that (i) there are only two states, say 0 and 1, and (ii) there is a symmetric rate of change between the two states 0 and 1, or, equivalently, both types of substitution change are equally likely on a given branch. We call the probability $\Pr[s_x = c \mid s_f = c]$ the **conservation probability** on that branch, denoted $p$ (by the symmetry assumption, $p$ is the same for $c=0$ or $c=1$). Throughout this paper, the accuracy of a reconstruction method is represented as an increasing function of the conservation probability rather than a decreasing function of the probability of change.

We analyze reconstruction accuracy in two evolutionary models: the equal branch length model (Oakley et al., 2005; Lee et al., 2006), which assumes that change happens mostly at speciation events (vertices) and therefore the length of each branch is irrelevant, and the distance model (Oakley et al., 2005; Lee et al., 2006), which assumes that mutations occurred continuously during the course of evolution and hence the branch length is no longer a constant. Both models have their advantages and disadvantages in ancestral state reconstruction (reviewed in Cunningham, 1999).

Under the symmetric evolutionary model given above, the ancestral state at the root of the phylogenetic tree begins with a character state and evolves with probability $1-p$ of change on the branches of the tree; hence the extant taxa would receive one of many possible distributions of character states 0 and 1. These states in extant taxa are the data used by the Fitch method (Fitch, 1971) to reconstruct the most parsimonious character state at the root as follows. This method assigns a set of states to each node one by one downward in the tree, starting with the leaves and using the subsets previously computed for the node's descendants. For each leaf node, the observed state forms the state subset. For the internal nodes, the following rule is used. Assume $A$ is an internal node with descendants $B$ and $C$. The state subset $S_A$ is calculated from the state subsets $S_B$ and $S_C$ by:

$$S_A = \begin{cases} S_B \cup S_C & \text{if} \quad S_B \cap S_C = \boldsymbol{f}, \\ S_B \cap S_C & \text{if} \quad S_B \cap S_C \neq \boldsymbol{f}. \end{cases}$$

The state subset at the root contains all the possible states that will be assigned to it. The method **unambiguously** reconstructs a state at the root if the state subset contains only that state and **ambiguously** reconstructs a state if the state subset contains both 0 and 1.

Since we are concentrating on state reconstruction in a symmetric evolutionary model with two states 0 and 1, the reconstruction's accuracy is independent of the prior distribution of the states at the root. Hence, for the Fitch method, the **unambiguous (reconstruction) accuracy** is:

$$UA = P_A[\{1\} \mid 1],$$

which is the probability that the state 1 is correctly reconstructed at the root $A$.

The **ambiguous (reconstruction) accuracy** is:

$$AA = P_A[\{0,1\} \mid 1].$$

The **reconstruction accuracy** of the method is:

$$RA = P_A[\{1\} \mid 1] + \frac{1}{2} P_A[\{0,1\} \mid 1],$$

where the second term in the expression $RA$ simply indicates that when either state (0 or 1) is equally parsimonious as a root state then we can select either state with equal probability.

**RESULTS**

We first consider the reconstruction accuracy of the Fitch method on the complete binary phylogenetic tree on $2^n$ taxa in the equal branch length model. As $n$ tends towards infinity, the unambiguous accuracy $UA_n(p)$ converges to 1/3 when $1/2 \leq p \leq 7/8$ and to

$$f(p) = \frac{1}{2(2p-1)}\left[(4p-3) + \frac{1}{2p-1}\sqrt{(8p-7)(4p-3)}\right]$$

when $p \geq 7/8$ as shown in (Steel, 1989). When $n$ approaches infinity, the conservation probability on any root-to-leaf path converges to 1/2. Therefore, when $1/2 \leq p \leq 7/8$, the unambiguous reconstruction accuracy of using the all the terminal taxa is smaller than 1/2, the limit of the conservation probability. As $n$ approaches infinity, the reconstruction accuracy $RA_n(p)$ of using all the terminal taxa converges to 1/2 when $p \leq 7/8$ and to

$$f(p) + \frac{1-p}{2p-1} = \frac{1}{2} + \frac{1}{2(2p-1)^2}\sqrt{(8p-7)(4p-3)} \geq \frac{1}{2}$$

when $p \geq 7/8$ as shown in (Steel, 1989). Hence, on a large, complete binary phylogenetic tree, the conservation probability on a root-to-leaf path is larger than the unambiguous accuracy when $p$ is small, but smaller than the reconstruction accuracy of using all the terminal taxa (see Figure 1).

Next, we consider the comb-shaped tree with $n$ leaves as shown in Figure 2. Note that in the equal branch length model, a descendant leaf of the root is closer in evolutionary distance to the root than other leaves in a larger clade. The unambiguous accuracy $UA_n(p)$ of using all the terminal taxa in the tree is:

$$a_1 l_1^n + a_2 l_2^n + \frac{p(2-3p+2p^2)}{2-3p^2+2p^3},$$

where $l$ indicates the roots of the characteristic equation:

$$l^2 + (1-p)l + p(1-3p+2p^2) = 0.$$

Since $|l_1|, |l_2| < 1$ for $0 < p < 1$, $UA_n(p)$ converges to $\frac{p(2-3p+2p^2)}{2-3p^2+2p^3}$ as $n$ tends towards infinity. The limit can be easily shown to be less than $p$. Similarly, the reconstruction accuracy $RA_n(p)$ converges to $\frac{1+2p-3p^2+2p^3}{2(2-3p^2+2p^3)}$. As shown in Figure 2, the limit of the reconstruction accuracy is also smaller than $p$, the conservation probability on the branch leading to the descendant leaf of the root.

The observation on the comb-shaped trees applies to any asymmetric phylogenetic trees $T$ in which a descendant leaf of the root $A$ is on a branch that is shorter than the branch leading to a large clade as illustrated in Figure 3. We now establish this result under a model in which the branch length is not constant. Let $Y$ be the descendant leaf and $Z$ the other non-descendant leaf of A. We assume that the conservation probability on the branches leading to $Y$ and $Z$ be $p_1$ and $p_2$ respectively, and set

$$P_Z[\{a\}|a] = \boldsymbol{b}, \quad a \in \{0,1\}$$

and

$$P_Z[\{b\}|a] = \boldsymbol{g}, \quad a \neq b \in \{0,1\}.$$

We have

$$P_A[\{0\}|0] = p_1(1 - p_2\boldsymbol{g} - (1-p_2)\boldsymbol{b})$$

and

$$P_A[\{0,1\}|0] = (p_1 p_2 + (1-p_1)(1-p_2))\boldsymbol{g} + (p_1(1-p_2) + (1-p_1)p_2)\boldsymbol{b}.$$

The accuracy of the Fitch method for reconstructing the root state is:

$$RA = P_A[\{0\}|\{0\}] + \frac{1}{2}P_A[\{0,1\}|0] = p_1 + \frac{1}{2}(1 - p_1 - p_2)\boldsymbol{g} + \frac{1}{2}(p_2 - p_1)\boldsymbol{b}.$$

When $p_1 \geq p_2 > \frac{1}{2}$, the reconstruction accuracy $RA$ is less than $p_1$ (since $\beta > 0$ and, in general, $\gamma > 0$). This shows that reconstructing the root state from all the leaf states in $T$ using maximum parsimony is less accurate than using just the state of a leaf adjacent to the root, whenever the branch leading to this leaf is not longer than the branch leading to the clade.

More interestingly, the reconstruction accuracy of the local or marginal ML method is just equal to $p_1$ even with multiple states. For simplicity, we show this only for two-state models as follows. When we say $D$ is a state configuration of the terminal taxa, we mean that $D$ contains a state for each terminal taxon in $T$. For a root state $c$ and a state configuration $D$ of the terminal taxa, we use $P(D|c)$ to denote the probability that $c$ evolves into the states specified by $D$ at the leaves. For any state configuration $D_Z$ of the terminal taxa below the node $Z$ and $s=0, 1,$ the term $sD_Z$ denotes the state configuration of all the terminal taxa in which $Y$ receives state $s$ and other taxa receive the states specified by $D_Z$. For any $D_Z$, we have

$$P(0D_Z | 0) = p_1(p_2 P(D_Z | 0) + (1 - p_2) P(D_Z | 1))$$

and

$$P(0D_Z | 1) = (1 - p_1)((1 - p_2) P(D_Z | 0) + p_2 P(D_Z | 1)).$$

Hence,

$$P(0D_z | 0) - P(0D_z | 1) = (p_1 + p_2 - 1) P(D_z | 0) + (p_1 - p_2) P(D_z | 1).$$

If $p_1 \geq p_2 > 1/2$, we have $P(0D_Z | 0) > P(0D_Z | 1)$ (since, in general, $P(D_Z|0)>0$). This implies that the local ML method infers 0 as the root state with the probability

$$\sum_{D_Z} P(0D_Z | 0) = p_1.$$

Notice that, in the two arguments we have presented above (for parsimony and marginal ML), we have imposed no assumption concerning the conservation probabilities on branches within the tree, other than (i) $p_1 \geq p_2 > \frac{1}{2}$ and (ii) the other conservation probabilities are non-degenerate (so $\gamma>0$ and $P(D_Z|0)>0$). We have shown that the accuracy of the Fitch method for reconstructing ancestral character states at the root from all terminal states in a phylogenetic tree can be smaller than the conservation probability on a path from the root to a nearest leaf. To find out how often this happens, we conducted a computer simulation test. We generated random phylogenetic trees using the Yule model. The generation procedure starts with a single root node. In each step, the procedure randomly selects one leaf with uniform distribution from the current tree and adds two descendants to it. The process terminates when the generated phylogeny has the required number of leaves.

For each random phylogenetic tree, we calculated and compared the conservation probability on the shortest root-to-leaf path, the conservation probability on the longest root-to-leaf path and the accuracy of reconstructing the ancestral state at the root from all the leaf states. We assumed that all branches had the same length and that the conservation probability is $p$. For $N = 9, 15, 20$ and $p = 0.5 + 0.01i$, $0 \leq i \leq 49$, we generated 5000 random phylogenetic trees with $N$ leaves and the conservation probability $p$ on each branch. The left panel of Figure 4 gives the number of generated phylogenetic trees in which the conservation probability on the shortest root-to-leaf path is larger than the accuracy of reconstructing the ancestral root state from all the leaf states with the Fitch method. When $p$ is in the range of 0.5 and 0.8, the conservation probability on the shortest root-to-leaf path is larger than the accuracy of reconstructing the correct state at the root in a large portion of trees. When $p$ exceeds 0.83, the number of 'bad' trees decreases rapidly. The right panel of Figure 4 shows the reconstruction accuracy of these three different reconstructions from some sampled trees. It is well known that the Yule model tends to produce trees that are, on average, more balanced than most real reconstructed trees (Aldous, 2001; Blum and Francois, 2006) and so we expect the level of support for the accuracy of root state reconstruction using a single species to be higher on real trees.

In general, the reconstruction accuracy of using a subset of the terminal taxa can also be higher than that obtained by using all the terminal taxa. For example, for the phylogenetic tree given in Figure 5 in which the conservation probability in each branch is 0.71, the reconstruction accuracy is 0.5878 if all the leaf states are used and 0.5916 if the states of only the four closest leaves indicated in the figure are used. This is true as long as the conservation probability is in the range from 0.5 to 0.82.

**DISCUSSION**

In studying how the accuracy of ancestral state reconstruction depends on taxon sampling, we demonstrated that more taxa are not necessarily better for ancestral state reconstruction with the Fitch method under the assumption that the true phylogenetic tree was given. This also happens with the maximum likelihood method. Our results and analyses have several implications.

First, taxon sampling has a subtle effect on the accuracy of ancestral state reconstruction. Unambiguous and ambiguous accuracy are considered separately by Salisbury and Kim. Our analyses indicate that unambiguous and ambiguous accuracy first decrease and then increase with the number of taxa sampled in reconstructing the root state in a phylogenetic tree. This pattern of increased accuracy with a large dataset of sampled taxa is consistent with the simulation results in Salisbury and Kim (2001). In our paper, we define the reconstruction accuracy to be the unambiguous accuracy plus half of the ambiguous accuracy for the Fitch method. The reconstruction accuracy is much more sensitive to the tree structure and does not monotonically depend on the size of taxon sampling especially when the given phylogenetic tree is asymmetric. As a result, researchers may need to decide how to select taxa from the observable extant species in reconstructing the root state of a clade. In certain cases, a single extant taxon at the end of a slowly evolving lineage (basal to the root) may provide a more accurate estimate of the root state than a tree-based analysis involving all the taxa (in contrast to Crisp and Cook (2005)).

Secondly, both the parsimony and ML methods attempt to incorporate the tree structure into ancestral state reconstruction. Suppose that 88 lineages formed a very recently diversified clade with a very long stem and a single sister lineage. If the 88 lineages have state 1, but the sister lineage has state 0, which state should their common ancestor have? This is exactly the situation when both fossil record and extant data are used for ancestor state reconstruction (e.g. evolution of body size in the Caniformia in Finarelli and Flynn (2006)). This is also the scenario when an outgroup is used in reconstructing the ancestral state. Our analysis concludes that 0 is selected as the root state by the Fitch method and the local ML method. Hence, when fossil record is used, it is very likely for the reconstructed ancestor to take the fossil state. This suggests that the Fitch and even local ML methods might not explore the full power of incorporating the fossil record into extant data. Therefore, when both fossil record and extent data are available, one may need to select an optimal subset of taxa carefully or to apply another sophisticated method for the reconstructing ancestor state. It also suggests that caution should be used in drawing conclusions on testing evolutionary hypotheses with the ancestral state reconstruction approach.

Finally, we have derived counterintuitive phenomena under a particularly simple evolutionary model, namely, the two-state symmetric model where branch lengths can be equal or unequal. A natural question that empiricists may ask is how often this counterintuitive situation arises in practice. To take one example, for certain data, branch lengths might be expected to satisfy a molecular clock – this amounts to allowing each edge $e$ to have its own conservation probability $p(e)$ but requiring that the sum of – log($2p(e)$-1) is to be constant on each root-to-leaf path. A phylogenetic tree with this clock constraint is said to be ultrametric. For an ultrametric tree, it might be expected that using all the taxa results in more accurate root state estimation than using a subset of the taxa only. The simulations of Salisbury and Kim (2001) and Zhang and Nei (1997) suggested that this is often the case. Salisbury and Kim investigated how the accuracy of reconstructing root states responds to size changes in taxon sampling in an ultrametric phylogeny generated in a Yule model. Their results indicate that reconstruction accuracy is generally increased by using more taxa.

However, once again, we find this general trend is not universally valid. More precisely, our simulation test shows that even with an ultrametric phylogenetic tree, the Fitch method or the joint ML method using a particular subset of terminal taxa can be more accurate (or at least as accurate) for ancestral state reconstruction than using all terminal taxa. We also observed that this holds for the four-state symmetric model. A related phenomenon was shown by Mossel (2001) for a certain asymmetric model using information-theoretic methods. In summary, the phenomena we have described is not restricted to trees that have a highly unusual set of (non-clocklike) branch lengths.

Despite this further counterintuitive result, we end by offering the following positive conjecture: for any ultrametric phylogenetic tree and a symmetric model, the Fitch parsimony method using all terminal taxa is more accurate (or at least as accurate) for ancestral state reconstruction than using any particular terminal taxon. Note that all root-to-leaf paths have the same conservation probability under a clock model.

**ACKNOWLEDGMENTS**

The authors thank the anonymous reviewers and Todd Oakley for pointing out related references and helpful comments on the relevance of our results to applying the fossil records to ancestral state reconstruction. L.X. Zhang gratefully acknowledges the NUS ARF grant R-146-000-068-112 and NSFChina3052802 for partially supporting this project. G.L. Li was partially supported by the NUS President's Graduate Fellowship. L.X. Zhang also thanks Webb Miller for stimulating this research by pointing out the paper by Lucena and Haussler to him.

**Figure captions:**

**Figure 1.** The accuracy of reconstructing the ancestral state at the root from all the leaf states in the complete phylogenetic tree on $2^n$ taxa when $n$ is large. *UA* denotes the unambiguous reconstruction accuracy; *RA* denotes the reconstruction accuracy.

**Figure 2.** Comb-shaped phylogenetic trees and the accuracy of reconstructing the ancestral state at the root from all terminal taxa in the limit case. *UA* denotes the unambiguous reconstruction accuracy; *RA* denotes the reconstruction accuracy.

**Figure 3.** Imbalanced trees in which there is a large clade and a sister lineage.

**Figure 4.** The left part gives the number of the random phylogenetic trees in which the accuracy of reconstructing the ancestral state at the root from all the leaf states is lower than the conservation probability on the shortest root-to-leaf path. The right part shows the accuracy of different reconstructions: the squares represent the reconstruction accuracy of using all the leaf states; plus signs and stars represent the conservation probability on the shortest and longest root-to-leaf paths, respectively.

**Figure 5.** In this phylogenetic tree, the Fitch method reconstructs the correct state at the root using all terminal taxa with less accuracy than using just the states of the four closest terminal taxa.

**Figure 1:**

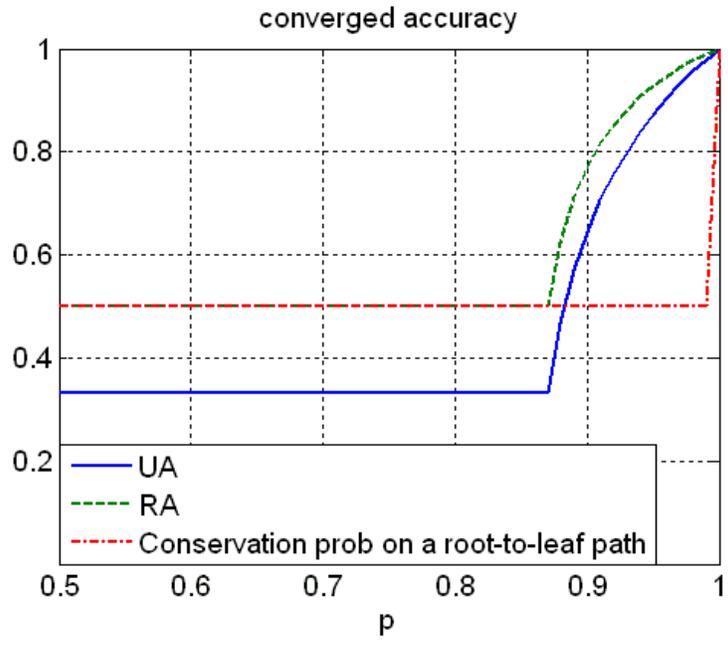

**Figure 2:**

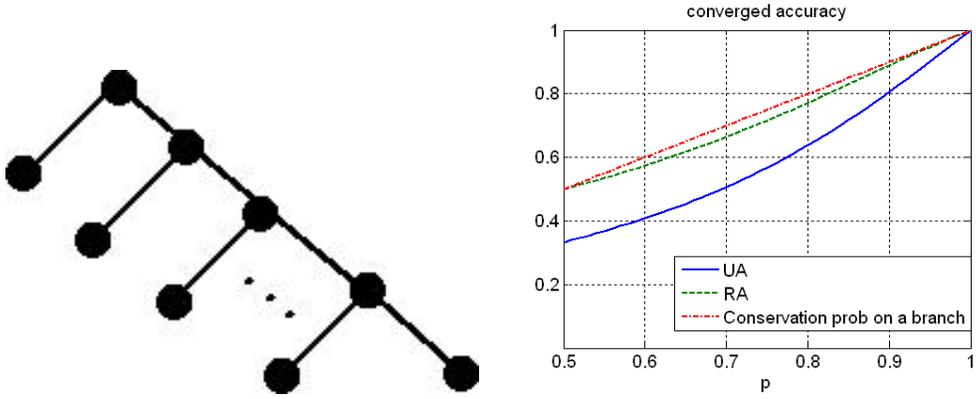

**Figure 3:**

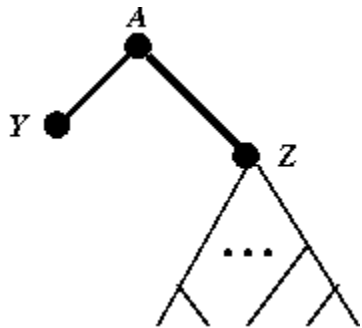

**Figure 4.**

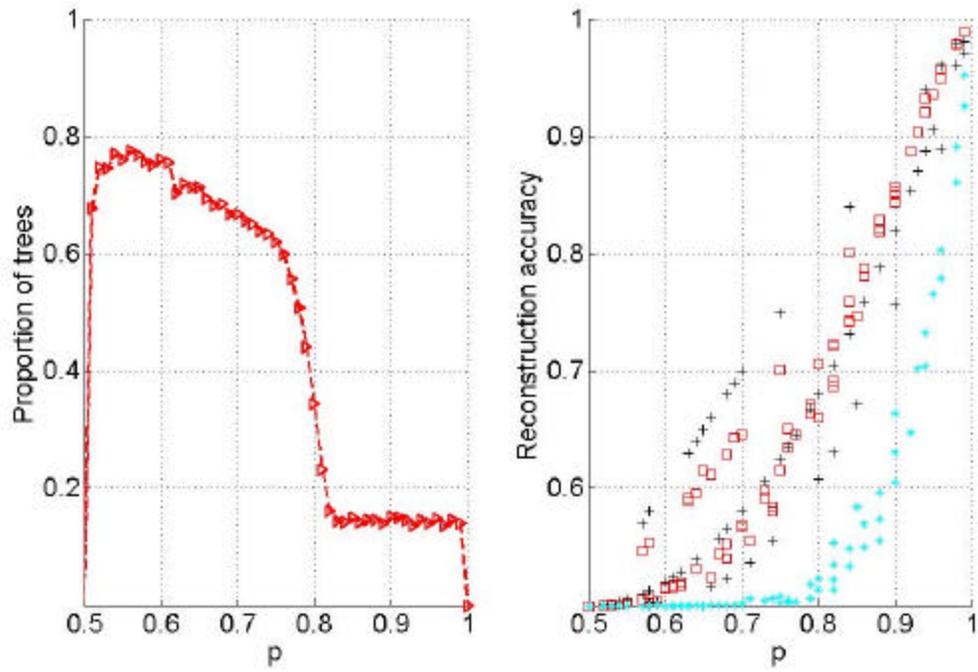

**Figure 5.**

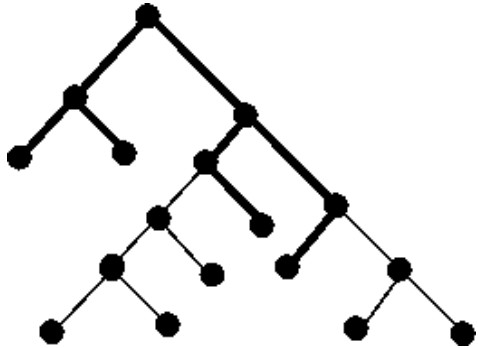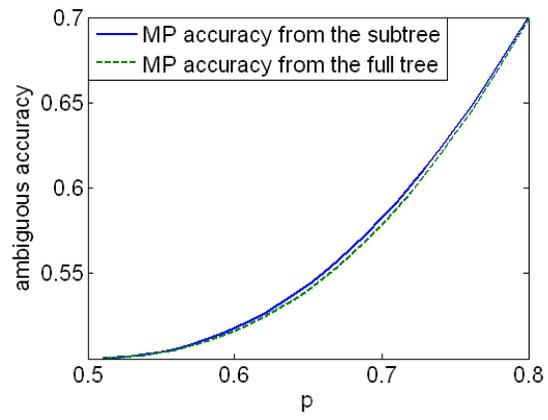